# Evolution of Maximum Bending Strain on Poisson's Ratio Distribution

Yang Li,[1,†] Le Zhang,[2,†] Dehua Wang,[1] Limei Hou,[1] Shanmei Du,[1] Yang Deng,[1] Yanfeng Du,[1] Yingfei Xin,[1] Chongyang Fu,[1] Yan Gu[3,*] & Xiaoxiong Wang[1,4,5,6,*]

**AFFILIATIONS**

[1]College of Physics Science, Qingdao University, Qingdao 266071, China.
[2]School of Physics and Electronic Engineering, Jiangsu Normal University, Jiangsu 221116, China.
[3]School of Mathematics and Statistics, Qingdao University, Qingdao 266071, China.
[4]Collaborative Innovation Center for Eco-Textiles of Shandong Province, and State Key Laboratory of Bio-Fibers and Eco-Textiles, Qingdao University, Qingdao 266071, China.
[5]University-Industry Joint Center for Ocean Observation and Broadband Communication, College of Physics, Qingdao University, Qingdao 266071, China.
[6]Weihai Innovation Research Institute of Qingdao University, Weihai 264200, China.
[†]These authors contributed equally to this work.
[*]Corresponding author.
e-mail: guyan1913@163.com; wangxiaoxiong@qdu.edu.cn

In recent years, new flexible functional materials have attracted increasing interest, but there is a lack of the designing mechanisms of flexibility design with superstructures. In traditional engineering mechanics, the maximum bending strain (MBS) was considered universal for describing the bendable properties of a given material, leading to the universal designing method of lowering the dimension such as thin membranes designed flexible functional materials. In this work, the MBS was found only applicable for materials with uniformly distributed Poisson's ratio, while the MBS increases with the thickness of the given material in case there is a variation Poisson's ratio in different areas. This means the MBS can be enhanced by certain Poisson's ratio design in the future to achieve better flexibility of thick materials. Here, the inorganic freestanding nanofiber membranes, which have a nonconstant Poisson's ratio response on stress/strain for creating nonuniformly distributed Poisson's ratio were proven applicable for designing larger MBS and lower Young's modulus for thicker samples.

In the past few years, the development of flexible materials has gradually become the frontier of electronic technology and functional materials [1,2,3,4]. Compared with traditional materials, flexible materials have been widely studied and applied in the fields of electronic skin, electronic medical devices and light-emitting devices due to their lightweight, high portability and good biocompatibility [5,6,7,8,9,10,11,12]. However, due to the inevitable nonuniformity, the flexible material composed of a single material can easy cause material changes under bending, such as the nonuniform Poisson's ratio effect, which in turn affects the mechanical properties of the material [13,14]. At present, the research on the mechanical properties of materials is more focused on the flexible design of materials, and the mechanical properties under the nonuniform Poisson's ratio effect are relatively rare.

Basically, two ways are used for designing new systems of flexible functional materials including structural design based on mechanics principle and properties design based on flexible materials like organics. The mechanical design is an emerging area for creating inorganic systems. The construction scheme of material flexibility includes the structural design method represented by dimension decrease, structure design [15,16,17], and the chemical design method by changing the strength of chemical coordination bond [18, 19], and molecular chain structure [20, 21]. The dimension decrease strategy has been widely applied in the field of dielectric materials design. Here the most important evaluation indicator of material flexibility in dimension reduction is the MBS [22, 23]. The equation used to calculate the MBS is $\varepsilon_{max} = h/D_{max}$ [24, 25], where $h$ is the thickness of a membrane and $D_{max}$ is the diameter of the curvature circle when the membrane is bent approaching crack. For example, in 2017, Zhou et al. prepared a two-dimensional (2D) Pb(Zr,Ti)O$_3$ membranes with the MBS of 1% on a mica sheet by epitaxial growth method and the membranes thickness was 50 nm [26]. Liu et al. prepared a 2D BaTiO$_3$ (BTO) membranes with 10.1% MBS on the Sr$_3$Al$_2$O$_6$ (SAO) sacrificial layer by pulsed laser deposition in 2019 and the membranes thickness was 120 nm [27]. Nowadays, the strategy expands its application area. In 2021, Tong et al. prepared one-dimensional (1D) flexible ice microfibers with 10.9% MBS by electric field enhanced growth and the nanofiber diameter was 4.3 μm [28]. These were micro examples. However, with the recent research discovery, some macroscopic samples with lower dimensional microstructures can also achieve such designs. In 2021, Ding et al. prepared flexible TiO$_2$ nanofibers and TiO$_2$ freestanding nanofiber membranes [29], and prepared Nb$_2$O$_5$ freestanding nanofiber membranes with photocatalytic properties in 2022 [30]. Zheng et al. prepared a flexible liquid metal freestanding nanofiber membranes with high conductivity in 2021 [31]. Wang et al. prepared inorganic flexible freestanding nanofiber membranes including (Ba,Ca)TiO$_3$:Pr$^{3+}$ and CaYAl$_3$O$_7$:Eu$^{3+}$ with photoluminescence properties in 2022 [32, 33].

What's more, the mechanical state is closely coupled to other properties. Dai et al.

prepared and measured the interface resistance change rate of LaAlO$_3$/SrTiO$_3$ (LAO/STO) freestanding membranes in 2019 and found that as the strain gradient of the material increases, the resistance change rate increases [34]. Pan et al. prepared freestanding BiFeO$_3$ membranes and STO membranes on SAO sacrificial layers in 2022. It was found that when the strain gradient of the material gradually increased, the polarization intensity also increased [35]. These results show the novel coupling of strain and physical parameters. Here, by demonstrating the MBS variation without changing the material and microstructure, we show that the strain state is more complex than commonly regarded. However, the mechanism is demonstrated as applicable to designing better MBS.

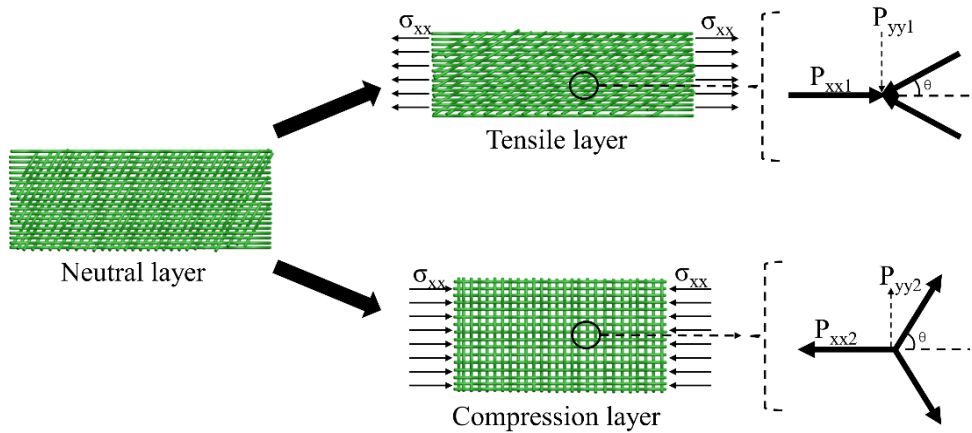

**Fig. 1. Diagram of Poisson's ratio of nanofiber layer changing with load.**

## Poisson's ratio distribution of nanofiber layer

During the high temperature sintering process, the organic impurities in the nanofiber material are easily decomposed and volatilized by heat to form a porous structure (Supplemental Fig. S1). As shown in Fig. 1, assume that the applied stress of the nanofiber tensile and compressive layers is consistent, when the nanofiber layer is in a stable state, the longitudinal area of the nanofiber stretch layer is smaller than that of the nanofiber compression layer ($S_{xx1} < S_{xx2}$), which makes the longitudinal load of the tensile layer smaller than the longitudinal load of the nanofiber compression layer ($P_{xx1} < P_{xx2}$).

$$P_{yy} = \frac{P_{xx}}{2} tan\theta \quad (1)$$

When the nanofiber layer is in a tensile state, the pores between the nanofiber are compressed, making the angle between the nanofiber smaller. When the nanofiber layer is compressed, the pores between the nanofiber expand, making the angle between the nanofibers larger. According to Eq. (1), the transverse load of a compressive layer is larger than that of a tensile layer ($P_{yy1} < P_{yy2}$). Since the transverse area of the nanofiber tensile layer is larger than that of the nanofiber compressive layer, the transverse stress of the tensile layer is smaller than that of the compressive layer ($\sigma_{yy1} < \sigma_{yy2}$). According to Hooke's law ($\sigma = E\varepsilon$), the transverse strain of the tensile layer is smaller than that of the compressive layer, and the longitudinal strain of tensile layer is equal to that of compressive layer. ($|\varepsilon_{yy1}| < |\varepsilon_{yy2}|, |\varepsilon_{xx1}| = |\varepsilon_{xx2}|$). According to the Poisson's ratio calculation equation ($v = -\varepsilon_{yy}/\varepsilon_{xx}$),

$$v_1 < v_2 \quad (2)$$

That is, the Poisson's ratio of the nanofiber compression layer is greater than that of the tensile layer.

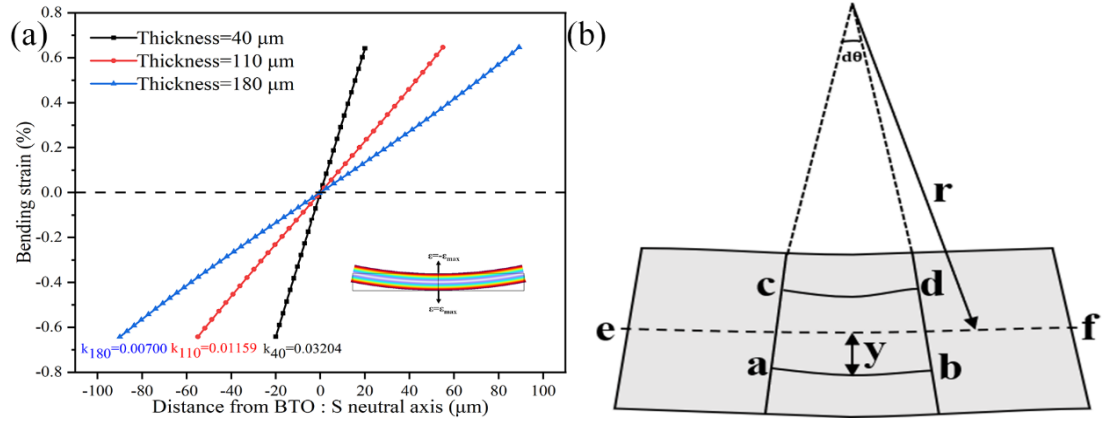

**Fig. 2. The change of internal bending strain of freestanding nanofiber membranes via finite element analysis (FEA). (a)** Strain variation at different positions of freestanding nanofiber membranes under fixed freestanding nanofiber membranes thickness. **(b)** The bending strain distribution diagram of each point on the cross section of the freestanding nanofiber membranes during the bending process.

## Bending strain in freestanding nanofiber membranes

In the FEA, the Young's modulus of the fixed freestanding nanofiber membranes is 61 MPa [36]. Since the Poisson's ratio of the material varies between 0.2 and 0.25 in the BTO ceramic material, we consider taking the neutral axis as the boundary, the Poisson's ratio of the compression layer from the supported freestanding nanofiber membranes is 0.25, and the Poisson's ratio of the tensile layer is 0.2 [37]. The roller support is given in the middle of the membranes, and the point is fixed at the roller support. Then a fixed bending normal stress increasing on both sides of the neutral axis is applied on both sides of the membranes. By changing the thickness of the freestanding nanofiber membranes, the radius of curvature of the membranes at each thickness is obtained. Then the bending strain at different positions in the membranes at different thicknesses is obtained, as shown in Fig. 2(a). The lower right illustration shows the bending strain distribution of the freestanding nanofiber membranes, which is the largest in the compressive and tensile layers of the freestanding nanofiber membranes. The calculation formula of bending strain can be derived from the bending strain distribution of each point on the cross-section of the freestanding nanofiber membranes in Fig. 2(b), where a-b and c-d are vertical lines, a-c and b-d are horizontal lines. When the freestanding nanofiber membranes are bent, tensile deformation occurs in the a-b longitudinal layer, and compressive deformation occurs in the c-d longitudinal layer. At the junction of the tensile and compressive layers, there exists a neutral axis e-f that is neither tensile nor compressive, and the stress at each point on the neutral axis is zero. Suppose the radius of curvature of the cross-section after bending is $r$, the angle of a-b with respect to the center of curvature is $d\theta$, the length of a-b before bending is $rd\theta$, and the length after bending is $(r+y)d\theta$, where $y$ is the distance of the a-b layer from the neutral axis, thus the bending strain $\varepsilon = [(r+y)d\theta - rd\theta]/rd\theta = y/r$, when $y = h/2$, the bending strain reaches the maximum value $\varepsilon_{\max} = h/D_{max}$, $D_{max}$ is the maximum curvature diameter of the cross section of the freestanding nanofiber membranes.

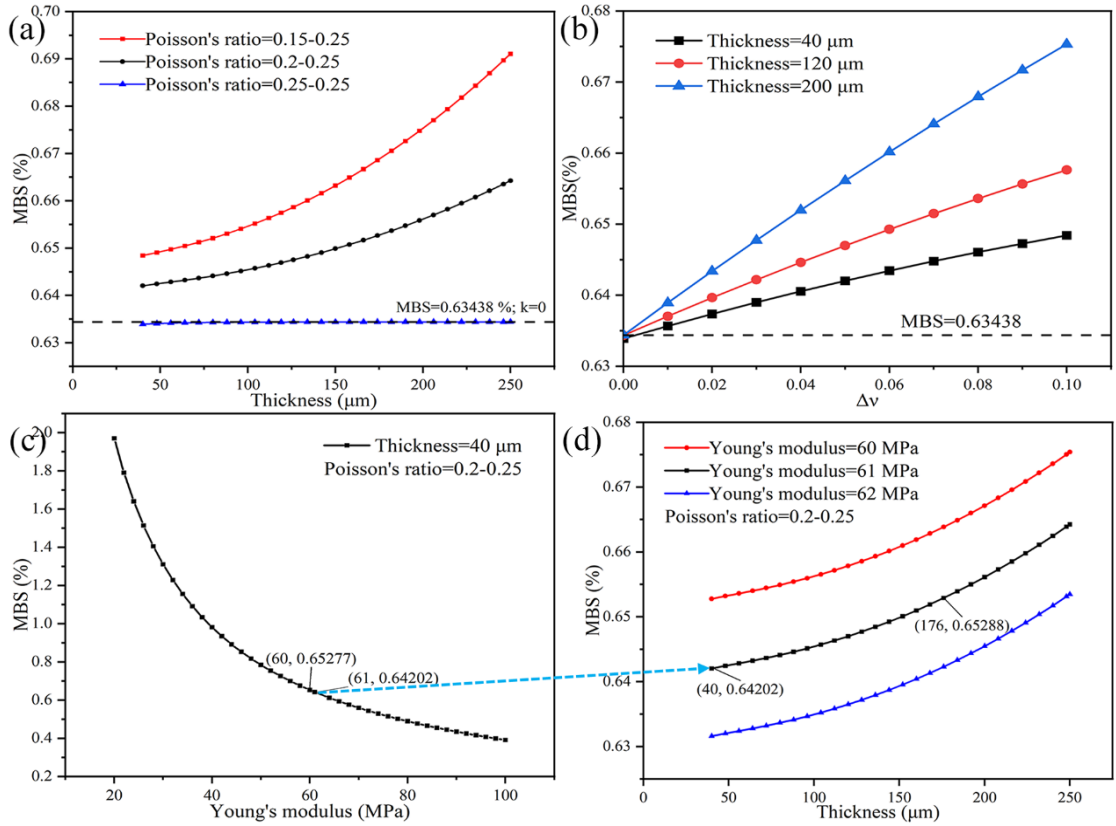

**Fig. 3. The MBS of freestanding nanofiber membranes varies with external and internal properties. (a)** The change of MBS with thickness under different Poisson's ratio distribution of freestanding nanofiber membranes. **(b)** The change of MBS with Poisson's ratio difference ($\Delta v$) under different freestanding nanofiber membranes thickness. **(c)** The variation of MBS with Young's modulus under the thickness of 40 μm freestanding nanofiber membranes. **(d)** The variation of MBS with thickness under different Young's modulus of freestanding nanofiber membranes.

**The effect of freestanding nanofiber membranes properties on MBS**

Figure 3(a) shows the change of MBS with thickness under different Poisson's ratio of the nanofiber layer [13]. It can be seen that the MBS is highly nonlinear, and as $\Delta v$ between the nanofiber layers decreases, the change of the MBS slows down. When the Poisson's ratio difference $\Delta v = 0$, the nonuniform material is transformed into a traditional uniform material, and the MBS of the membranes does not change with the thickness. By comparing the MBS-$\Delta v$ curves of freestanding nanofiber membranes with different thicknesses of freestanding nanofiber membranes, we obtained the equivalent MBS of the freestanding nanofiber membranes. When $\Delta v = 0$, MBS is a constant A controlled by material properties. When $\Delta v \neq 0$, the MBS of the freestanding nanofiber membranes varies with the membranes thickness, where $p_1$, $p_2$, $p_3$, and $p_4$ are polynomial coefficient affected by $\Delta v$.

$$\varepsilon_{max} = A, \quad \Delta v = 0, \tag{3}$$

$$\varepsilon_{max} = A(p_1 + p_2 h + p_3 h^2 + p_4 h^3), \quad \Delta v \neq 0. \tag{4}$$

According to the existing research, with the increase of sintering temperature, the grain size of the nanofiber surface increases, and then Young's modulus of the freestanding nanofiber membranes increases, and Young's modulus of the freestanding nanofiber membranes is positively correlated with the sintering temperature [36, 38]. As shown in Fig. 3(c), it can be seen that when the Young's modulus of the freestanding nanofiber membranes increases, the MBS of the freestanding nanofiber membranes decreases and the flexibility decreases. When the Young's modulus of the freestanding nanofiber membranes is fixed at 61 MPa, the MBS of the freestanding nanofiber membranes increases from 0.64202% at 40 μm to 0.65288% at 176 μm with the increase of the thickness of the freestanding nanofiber membranes, which exceeds the MBS of the freestanding nanofiber membranes at Young's modulus of 60 MPa and thickness of 40 μm. The results show that although the MBS of the freestanding nanofiber membranes decreases with the increase of the sintering temperature and becomes less flexible, the effect of increasing the sintering temperature on the flexibility of the freestanding nanofiber membranes can be changed by increasing the thickness of the freestanding nanofiber membranes.

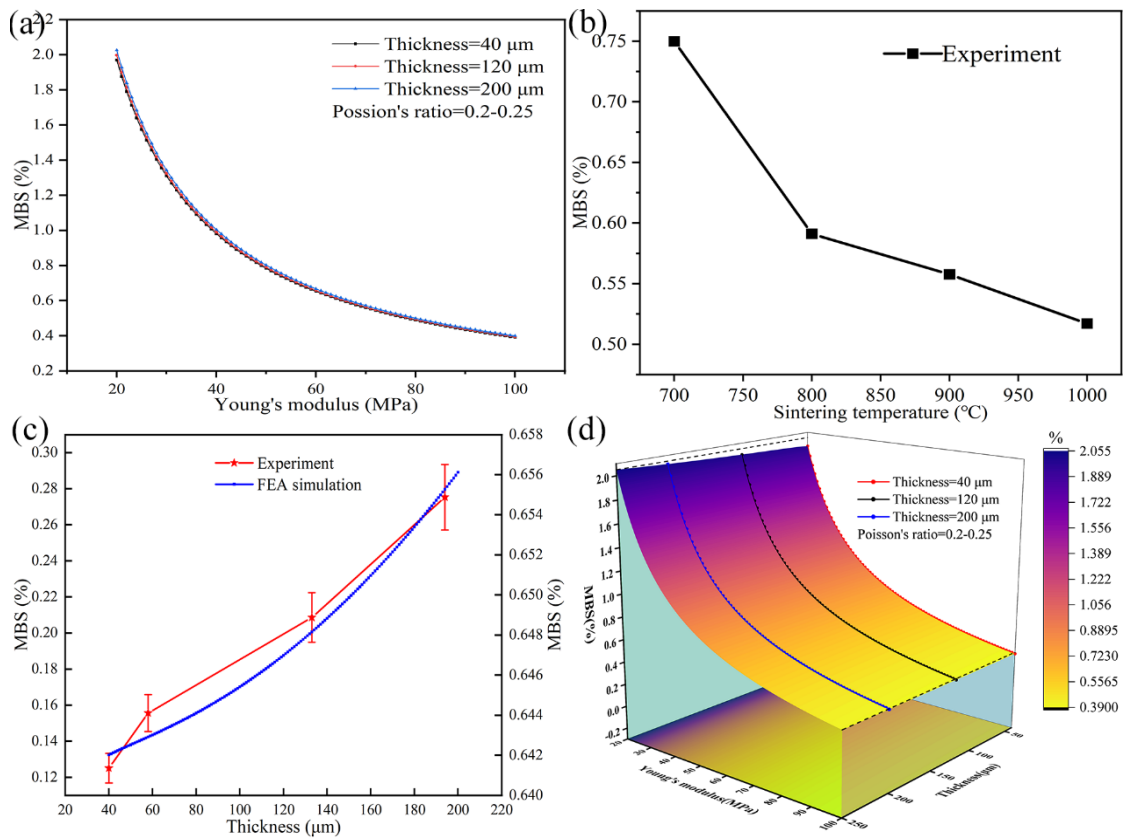

**Fig. 4. The MBS of ceramic-based freestanding nanofiber membranes varies with thickness and Young's modulus. (a)** The FEA of MBS with Young's modulus under freestanding nanofiber membranes with different thicknesses. **(b)** Experimental changes of freestanding nanofiber membranes MBS with sintering temperature. **(c)** The variation of MBS with the thickness of freestanding nanofiber membranes was obtained by FEA and experiment. **(d)** Three-dimensional (3D) FEA diagram of freestanding nanofiber membranes MBS with thickness and Young's modulus.

**The MBS experiment of freestanding nanofiber membranes**

After determining the influence of thickness on the material MBS, we designed some novel experimental results through this mechanism. For example, the MBS of ceramic-based freestanding nanofiber membranes at different sintering temperatures and thicknesses was obtained by using a three-point bending strain test system. As shown in Fig. 4(b), it can be seen that the MBS of the freestanding nanofiber membranes decreases with the increase of the sintering temperature, which is consistent with the trend of FEA in Fig. 4(a). The red line shown in Fig. 4(c) is the MBS curve of the freestanding nanofiber membranes obtained by the experiment with the thickness. The blue line is the MBS curve of the freestanding nanofiber membranes obtained by FEA with the thickness. Both the experiment and FEA show that the MBS of the freestanding nanofiber membranes increases with the increase of the thickness of the freestanding nanofiber membranes. Compared with the data measured by the FEA results, the data showed a certain fluctuation, which may be due to the uneven surface distribution caused by the instability of the current during the preparation of the freestanding nanofiber membranes [39, 40]. In order to further verify the proposed nonuniform Poisson's ratio model, we obtained the functional relationship between the equivalent MBS and the thickness by fitting the MBS curve of the ceramic-based freestanding nanofiber membranes measured in the experiment. Where $h$ is the thickness of the freestanding nanofiber membranes, the highest order term of the equivalent MBS varying with thickness is a cubic relationship, where $p_5$, $p_6$, $p_7$, and $p_8$ are polynomial coefficients varying with material properties and geometric configurations.

$$\varepsilon_{max} = p_5 + p_6 h + p_7 h^2 + p_8 h^3. \tag{5}$$

Figure. 4(d) shows the 3D variation of MBS in different Young's modulus and thickness. Using this 3D diagram, it can be clearly seen that if the MBS of the nanofiber membranes is to be increased, it can be achieved by increasing the thickness of the nanofiber membranes or reducing the Young's modulus. However, the method of increasing flexibility by reducing the Young's modulus is easily limited because its minimum value is greater than zero, but it is easier to achieve by increasing the thickness by the current technology [41].

## Conclusions and outlook

To summarize, we demonstrated the effect of Poisson's ratio distribution on the material MBS by designing a ceramic-based freestanding nanofiber membranes with pores. MBS can be affected by many factors, such as the increase of Young's modulus will lead to the decrease of MBS, but based on the findings of this work, this flexible damage can be compensated by increasing thickness. This effect is caused by the nonuniform Poisson's ratio of the material. In addition, by fitting the FEA and experimental data, we obtained the nonlinear variation of the MBS of the freestanding nanofiber membranes with the thickness, which is different from the traditional understanding that the MBS is a constant. Freestanding nanofiber membranes have shown challenges to conventional flexible design in related fields. Although nanofibers can be regarded as low-dimensional systems, freestanding nanofiber membranes exhibit exceptionally good inorganic flexibility at the macro scale. This study reveals the origin of this abnormal flexibility, which lays a theoretical foundation for the design of future metamaterials.

## Methods

### Sample preparation

Firstly, the precursor solution was synthesized by the sol-gel method, and then it was prepared through electrospinning to create nanofiber membranes. Finally, the nanofiber membranes were sintered by a muffle furnace to form freestanding nanofiber membranes (Supplementary Fig. S1 explain the details). The preparation material of the precursor solution is barium acetate ($C_4H_6BaO_4, \geq 99.00\%$). Tetrabutyl titanate ($C_{16}H_{36}O_4Ti, \geq 99.00\%$). Samarium (iii) nitrate · hexahydrate ($SmN_3O_9 \cdot 6H_2O, Mw = 444.47$). Glacial acetic acid ($C_2H_4O_2, 99.5\%$). Polyvinylpyrrolidone (PVP, Mw~1300000). First, a certain amount of deionized water is added to the reagent bottle. Then glacial acetic acid was added. When mixed evenly, barium acetate was added to mix well. Then PVP was added and stirred until uniform, PVP was 5wt% of the total mass of the solution. When PVP is fully dissolved, tetrabutyl titanate is added to dissolve fully. Finally, Samarium (iii) nitrate · hexahydrate was added until it was completely dissolved to form a $BaTiO_3: 0.1\%Sm^{3+}$ precursor solution.

### Finite element analysis

In the FEA, the control equation used is $\rho(\partial v/\partial t) - \nabla \cdot S = F_v$, where $\rho$ is the density of the material, $v$ is the velocity of the material, $S$ is the stress tensor of the material, and $F_v$ is the volume force of the material. Since the force state of the material is steady and constant during the experiment ($\partial v/\partial t = 0$), and thus the control equation is $0 = F_v + \nabla \cdot S$. The intrinsic equation of the material is $S = C:\varepsilon$, where $C = C(E, v)$ is the elastic matrix of the material, associated with Young's modulus $E$ and Poisson's ratio $v$ of the material. ":" is the double point tensor product and $\varepsilon$ is the strain tensor of the material. The coordinate system of the model is chosen as the global coordinate system. The material parameters of the model are listed in Table I. The grid division is a free quadrilateral grid, and the model is a very fine model with a maximum cell size of 3.6 μm and a minimum cell size of 0.0072 μm (Supplementary Table I).

## Acknowledgements

This work was supported by the National Natural Science Foundation of China (12104249), the Youth Innovation Team Project of Shandong Provincial Education Department (2020KJN015, 2021KJ013), State Key Laboratory of Bio-Fibers and Eco-Textiles (Qingdao University), No. GZRC202011 & ZKT46.

## Author contributions

Yang Li contributed to the preparation of the draft as the first author and proceeded with the finite elements analysis. Le Zhang contributed to the preparation of the draft as the co-first authors and proceeded with the finite elements analysis. Dehua Wang, Limei Hou, Shanmei Du, Yang Deng, Yanfeng Du, Yingfei Xin and Chongyang Fu contributed to measurements and post-experiments for the supporting information. Yan Gu and Xiaoxiong Wang contributed as a supervisor for the research.

## Competing interests

The authors declare no competing interests.

## References

1. Wen X, Li D, Tan K, Deng Q, Shen S. Flexoelectret: An Electret with a Tunable Flexoelectriclike Response. *Physical Review Letters*. **122**, 14 (2019).
2. Matsuhisa N, Niu S, O'Neill SJK, Kang J, Ochiai Y, Katsumata T, et al. High-frequency and intrinsically stretchable polymer diodes. *Nature*. **600**, 7888 (2021).
3. Wang W, Wang S, Rastak R, Ochiai Y, Niu S, Jiang Y, et al. Strain-insensitive intrinsically stretchable transistors and circuits. *Nature Electronics*. **4**, 2 (2021).
4. Yan Z, Xu D, Lin Z, Wang P, Cao B, Ren H, et al. Highly stretchable van der Waals thin films for adaptable and breathable electronic membranes. *Science*. **375**, 6583 (2022).
5. Wang S, Xu J, Wang W, Wang G-JN, Rastak R, Molina-Lopez F, et al. Skin electronics from scalable fabrication of an intrinsically stretchable transistor array. *Nature*. **555**, 7694 (2018).
6. Yu X, Xie Z, Yu Y, Lee J, Vazquez-Guardado A, Luan H, et al. Skin-integrated wireless haptic interfaces for virtual and augmented reality. *Nature*. **575**, 7783 (2019).
7. Xu S, Jayaraman A, Rogers JA. Skin sensors are the future of health care. *Nature*. **571**, 7765 (2019).
8. Kim Y, Chortos A, Xu W, Liu Y, Oh JY, Son D, et al. A bioinspired flexible organic artificial afferent nerve. *Science*. **360**, 6392 (2018).
9. Oh YS, Kim J-H, Xie Z, Cho S, Han H, Jeon SW, et al. Battery-free, wireless soft sensors for continuous multi-site measurements of pressure and temperature from patients at risk for pressure injuries. *Nature Communications*. **12**, 1 (2021).
10. Yang Q, Wei T, Yin RT, Wu M, Xu Y, Koo J, et al. Photocurable bioresorbable adhesives as functional interfaces between flexible bioelectronic devices and soft biological tissues. *Nature Materials*. **20**, 11 (2021).
11. Zhang Z, Wang W, Jiang Y, Wang Y-X, Wu Y, Lai J-C, et al. High-brightness all-polymer stretchable LED with charge-trapping dilution. *Nature*. **603**, 7902 (2022).
12. White MS, Kaltenbrunner M, Glowacki ED, Gutnichenko K, Kettlgruber G, Graz I, et al. Ultrathin, highly flexible and stretchable PLEDs. *Nature Photonics*. **7**, 10 (2013).
13. Du L, Luo S, Xu Y. Understanding nonlinear behaviors of auxetic foams using X-ray tomography and pore structure analysis. *Mechanics of Materials*. **165**, (2022).

14. Ma Y, Zheng Y, Meng H, Song W, Yao X, Lv H. Heterogeneous PVA hydrogels with micro-cells of both positive and negative Poisson's ratios. *Journal of the Mechanical Behavior of Biomedical Materials*. **23**, (2013).
15. Khang DY, Jiang HQ, Huang Y, Rogers JA. A stretchable form of single-crystal silicon for high-performance electronics on rubber substrates. *Science*. **311**, 5758 (2006).
16. Rogers JA, Someya T, Huang Y. Materials and Mechanics for Stretchable Electronics. *Science*. **327**, 5973 (2010).
17. Jang K-I, Chung HU, Xu S, Lee CH, Luan H, Jeong J, et al. Soft network composite materials with deterministic and bio-inspired designs. *Nature Communications*. **6**, (2015).
18. Zhang J, Liu G, Cui W, Ge Y, Du S, Gao Y, et al. Plastic deformation in silicon nitride ceramics via bond switching at coherent interfaces. *Science*. **378**, 6618 (2022).
19. Lai J-C, Jia X-Y, Wang D-P, Deng Y-B, Zheng P, Li C-H, et al. Thermodynamically stable whilst kinetically labile coordination bonds lead to strong and tough self-healing polymers. *Nature Communications*. **10**, 1 (2019).
20. Wang J, O'Connor TC, Grest GS, Ge T. Superstretchable Elastomer from Cross-linked Ring Polymers. *Physical Review Letters*. **128**, 23 (2022).
21. Jiang Y, Zhang Z, Wang Y-X, Li D, Coen C-T, Hwaun E, et al. Topological supramolecular network enabled high-conductivity, stretchable organic bioelectronics. *Science*. **375**, 6587 (2022).
22. Peng J, Snyder GJ. A figure of merit for flexibility. *Science*. **366**, 6466 (2019).
23. Peng J, Grayson M, Snyder GJ. What makes a material bendable? A thickness-dependent metric for bendability, malleability, ductility. *Matter*. **4**, 9 (2021).
24. Peng B, Peng R-C, Zhang Y-Q, Dong G, Zhou Z, Zhou Y, et al. Phase transition enhanced superior elasticity in freestanding single-crystalline multiferroic $BiFeO_3$ membranes. *Science Advances*. **6**, 34 (2020).
25. Siefert E, Cattaud N, Reyssat E, Roman B, Bico J. Stretch-Induced Bending of Soft Ribbed Strips. *Physical Review Letters*. **127**, 16 (2021).
26. Jiang J, Bitla Y, Huang C-W, Thi Hien D, Liu H-J, Hsieh Y-H, et al. Flexible ferroelectric element based on van der Waals heteroepitaxy. *Science Advances*. **3**, 6 (2017).
27. Dong G, Li S, Yao M, Zhou Z, Zhang Y-Q, Han X, et al. Super-elastic ferroelectric single-crystal membrane with continuous electric dipole rotation. *Science*. **366**, 6464 (2019).
28. Xu P, Cui B, Bu Y, Wang H, Guo X, Wang P, et al. Elastic ice microfibers. *Science*. **373**, 6551 (2021).
29. Zhang Y, Liu S, Yan J, Zhang X, Xia S, Zhao Y, et al. Superior Flexibility in Oxide Ceramic Crystal Nanofibers. *Advanced Materials*. **33**, 44 (2021).
30. Lin X, Xia S, Zhang L, Zhang Y, Sun S, Chen Y, et al. Fabrication of Flexible Mesoporous Black $Nb_2O_5$ Nanofiber Films for Visible-Light-Driven Photocatalytic $CO_2$ Reduction into $CH_4$. *Advanced Materials*. **34**, 16 (2022).
31. Ma Z, Huang Q, Xu Q, Zhuang Q, Zhao X, Yang Y, et al. Permeable superelastic

liquid-metal fibre mat enables biocompatible and monolithic stretchable electronics. *Nature Materials*. **20**, 6 (2021).
32. Dehua W, Longlong J, Yang Y, Ye L, Zifei M, Rui B, et al. Effect of grain size on macroscopic flexibility and luminescence intensity of inorganic (Ba,Ca)TiO3:Pr3+. *Journal of Alloys and Compounds*. **912**, (2022).
33. Wang D, Chen L, Jiang L, Yu Y, Lu Y, Li H, et al. A high temperature macroscopically flexible inorganic CaYAl3O7:Eu3+ nanofiber luminescent membrane. *Journal of Materials Chemistry C*. **10**, 19 (2022).
34. Zhang F, Lv P, Zhang Y, Huang S, Wong C-M, Yau H-M, et al. Modulating the Electrical Transport in the Two-Dimensional Electron Gas at LaAlO3/SrTiO3 Heterostructures by Interfacial Flexoelectricity. *Physical Review Letters*. **122**, 25 (2019).
35. Cai S, Lun Y, Ji D, Lv P, Han L, Guo C, et al. Enhanced polarization and abnormal flexural deformation in bent freestanding perovskite oxides. *Nature Communications*. **13**, 1 (2022).
36. Yan J, Han Y, Xia S, Wang X, Zhang Y, Yu J, et al. Polymer Template Synthesis of Flexible BaTiO3 Crystal Nanofibers. *Advanced Functional Materials*. **29**, 51 (2019).
37. Jia X, Zhang H-Q, Wang Z, Jiang C-L, Liu Q-J, Liu Z-T. Mechanical Properties, Born Effective Charge Tensors and High Frequency Dielectric Constants of the Eight Phases of BaTiO3. *Moscow University Physics Bulletin*. **72**, 4 (2017).
38. Ryou H, Drazin JW, Wahl KJ, Qadri SB, Gorzkowski EP, Feigelson BN, et al. Below the Hall-Petch Limit in Nanocrystalline Ceramics. *Acs Nano*. **12**, 4 (2018).
39. Korkut S, Saville DA, Aksay IA. Enhanced stability of electrohydrodynamic jets through gas ionization. *Physical Review Letters*. **100**, 3 (2008).
40. Beroz J, Hart AJ, Bush JWM. Stability Limit of Electrified Droplets. *Physical Review Letters*. **122**, 24 (2019).
41. Guo J, Fu S, Deng Y, Xu X, Laima S, Liu D, et al. Hypocrystalline ceramic aerogels for thermal insulation at extreme conditions. *Nature*. **606**, 7916 (2022).

— Supplementary Materials —

# Evolution of Maximum Bending Strain on Poisson's Ratio Distribution

Yang Li,[1,†] Le Zhang,[2,†] Dehua Wang,[1] Limei Hou,[1] Shanmei Du,[1] Yang Deng,[1] Yanfeng Du,[1] Yingfei Xin,[1] Chongyang Fu,[1] Yan Gu[3,*] & Xiaoxiong Wang[1,4,5,6,*]

**AFFILIATIONS**

[1]College of Physics Science, Qingdao University, Qingdao 266071, China.
[2]School of Physics and Electronic Engineering, Jiangsu Normal University, Jiangsu 221116, China.
[3]School of Mathematics and Statistics, Qingdao University, Qingdao 266071, China.
[4]Collaborative Innovation Center for Eco-Textiles of Shandong Province, and State Key Laboratory of Bio-Fibers and Eco-Textiles, Qingdao University, Qingdao 266071, China.
[5]University-Industry Joint Center for Ocean Observation and Broadband Communication, College of Physics, Qingdao University, Qingdao 266071, China.
[6]Weihai Innovation Research Institute of Qingdao University, Weihai 264200, China.
[†]These authors contributed equally to this work.
[*]Corresponding author.
e-mail: guyan1913@163.com; wangxiaoxiong@qdu.edu.cn

TABLE I. Material parameters used for FEA.

| Material | Young's Modulus (MPa) | Poisson's ratio | Density (Kg/m$^3$) | Thickness (μm) |
|---|---|---|---|---|
| BTO:S | 20-100 | 0.15-0.25 | 28 | 40-250 |

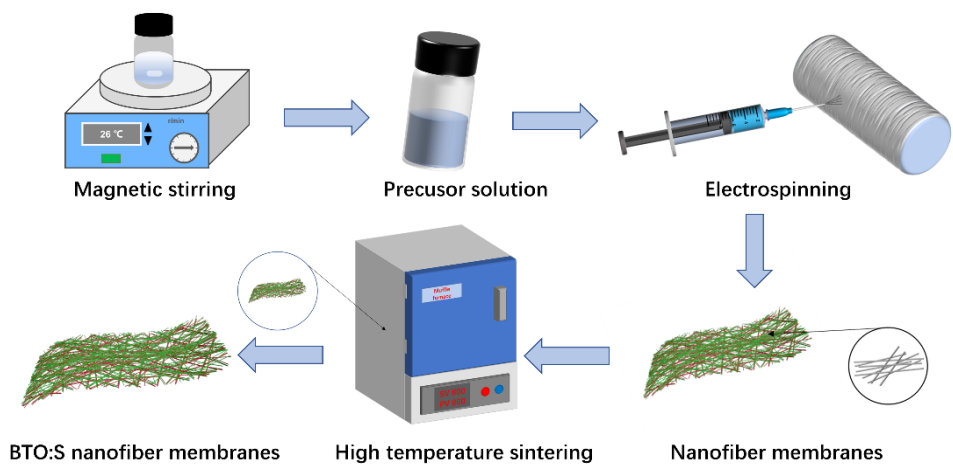

FIG. S1. Schematic diagram of electrospinning process and freestanding nanofiber membranes sintering.

Figure S1 shows a schematic diagram of the preparation of freestanding nanofiber membranes by the electrospinning method. The main components of the electrospinning device include an injector, a flat-tipped needle, a DC high-voltage source, and a drum. The gel-sol method is first configured with the inorganic $BaTiO_3:0.1\%Sm^{3+}$ (BTO:S) solution. Then the prepared inorganic BTO:S solution is filled in a syringe, after which a high-voltage power supply is connected to a flat-tipped needle. The drum is grounded, and the nanofibers generated on the flat head needle are adsorbed onto the drum through the action of positive and negative electric fields to form freestanding nanofiber membranes. The distance between the needle and the drum is 15 cm, the voltage applied to the flat-head needle is 18 KV, the speed of the solution in the syringe is 1.5 ml/h, and the speed of the drum is 300 $r \cdot min^{-1}$. The humidity of the environment is 20±5%, and the temperature is room temperature. After that, the obtained membranes were sintered in a muffle furnace for two hours and then cooled naturally. During the sintering process, the solvent in the solution evaporated, the decomposition of organic polyvinyl pyrrolidone by heat, and the inorganic BTO:S freestanding nanofiber membranes were finally formed.

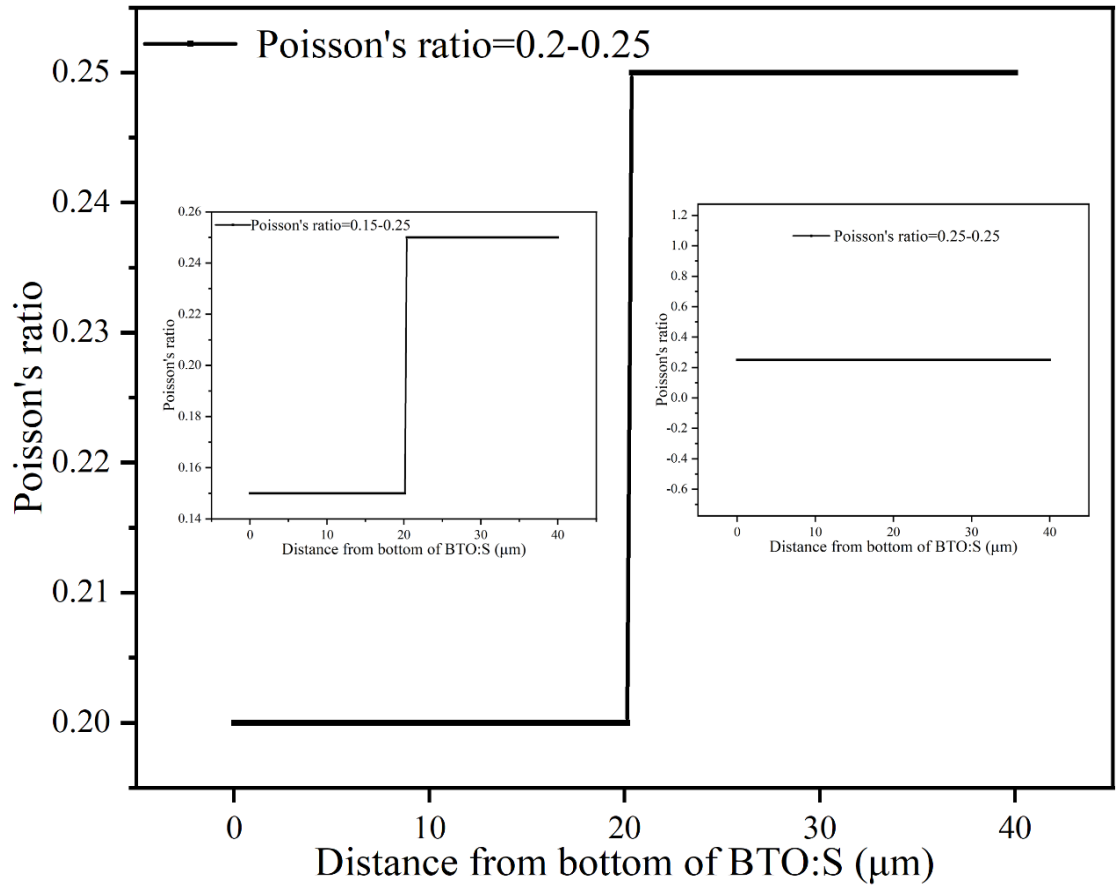

FIG. S2. Variation of Poisson's ratio with different membranes positions at a fixed freestanding nanofiber membranes thickness of 40 μm.

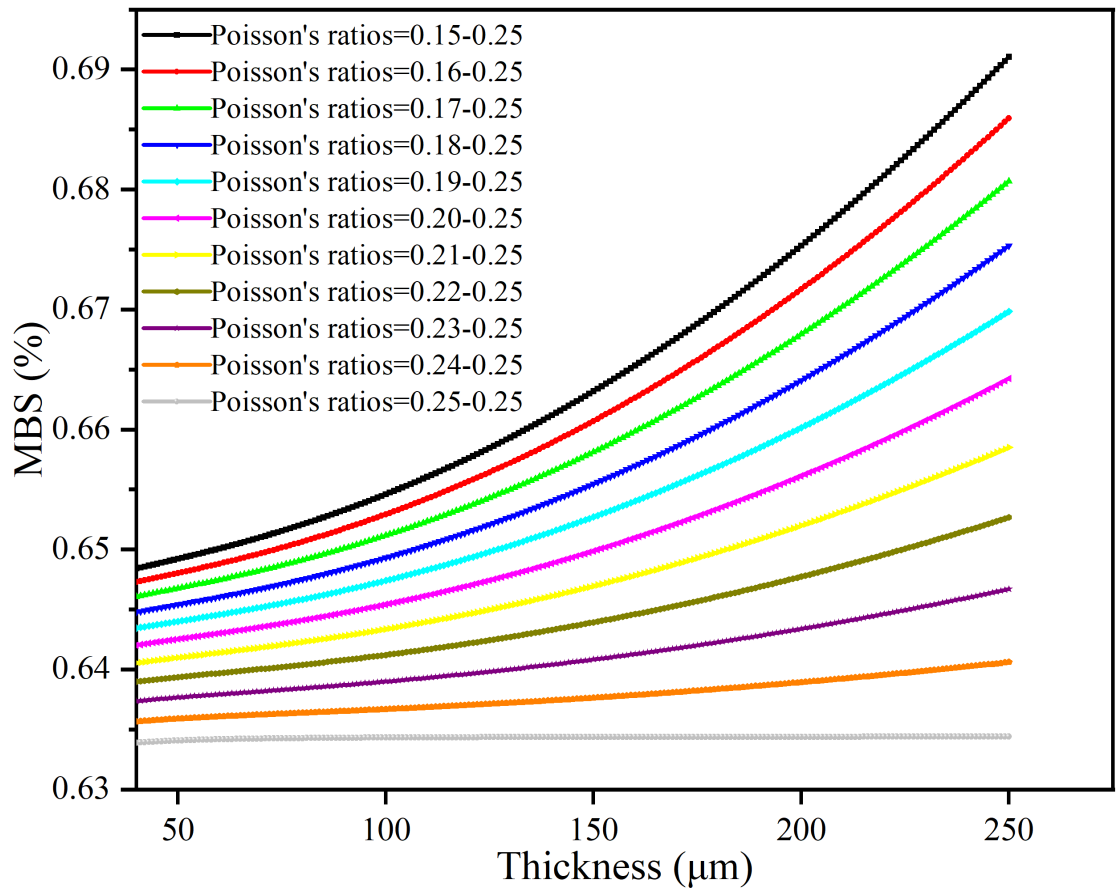

FIG. S3. Variation of MBS with thickness for different Poisson's ratio of nanofiber layers.

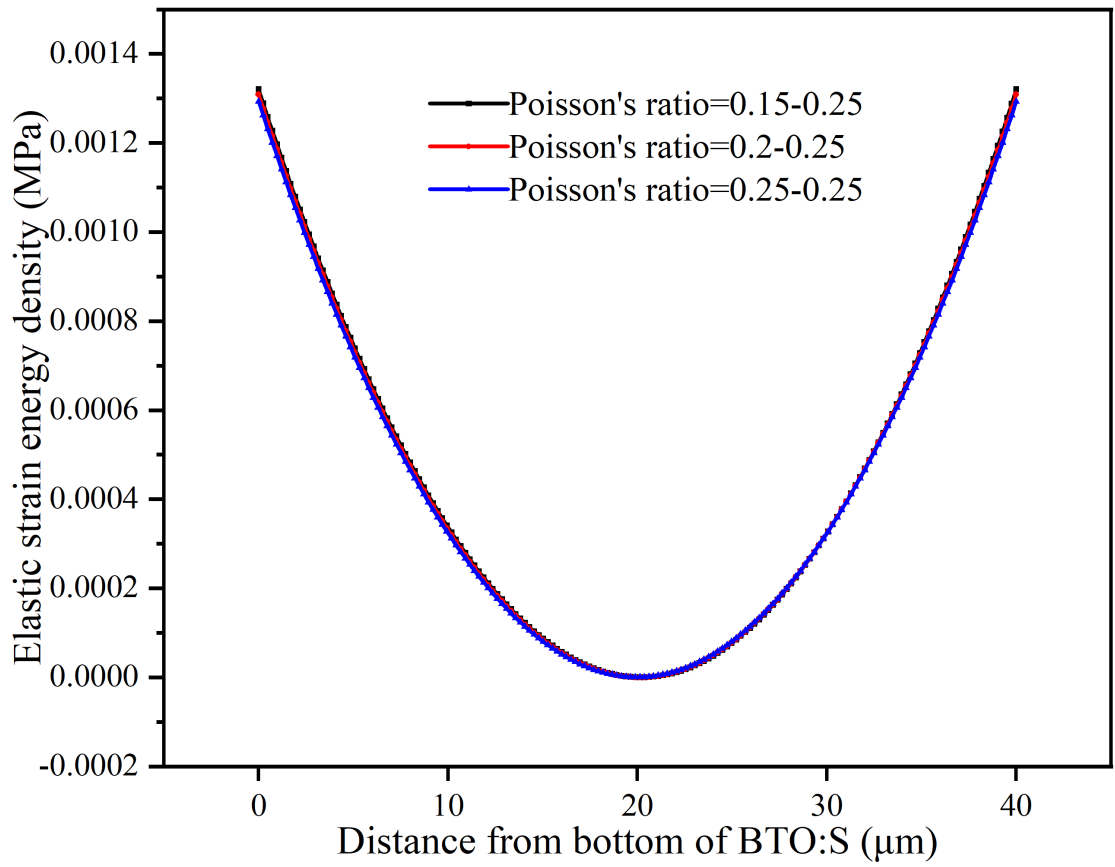

FIG. S4. Variation of elastic strain energy density with different membranes positions at different Poisson's ratio of nanofiber layers.

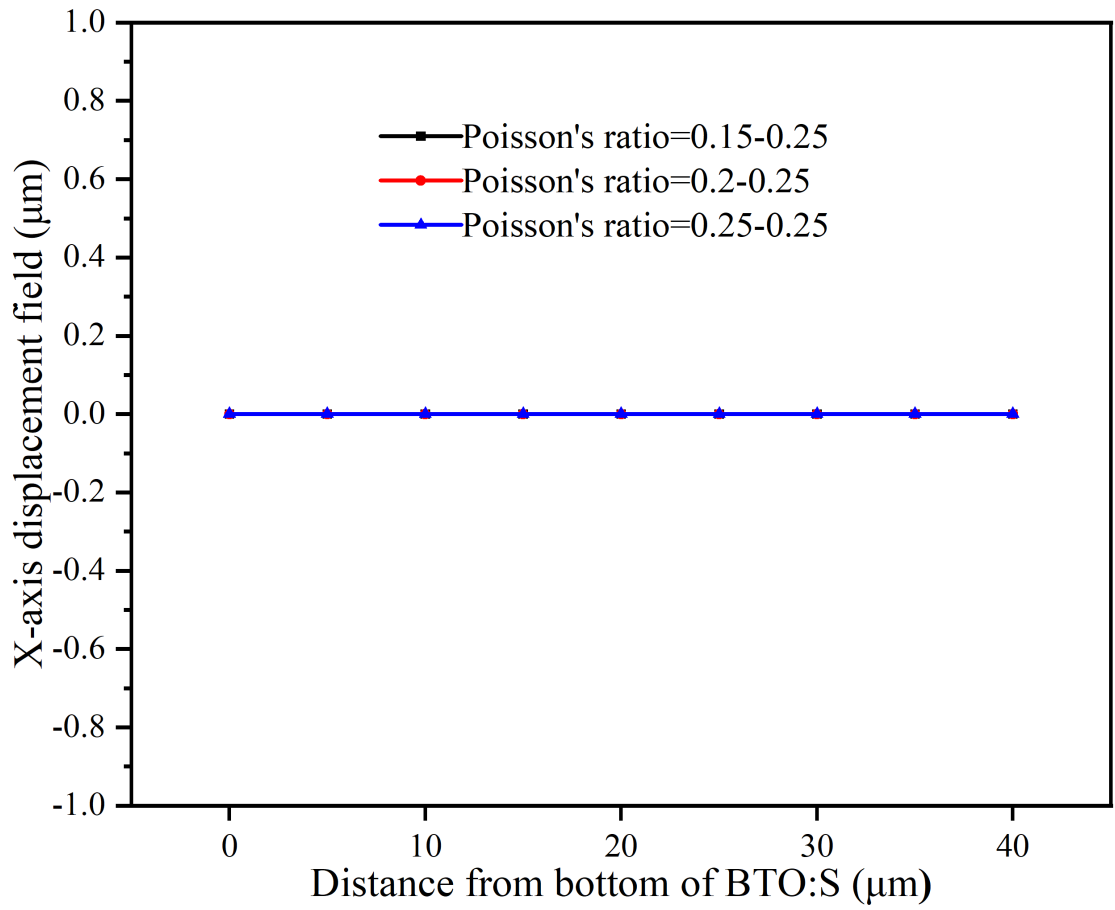

FIG. S5. Variation of X-axis displacement field components with different membranes positions for different Poisson's ratio of nanofiber layers.

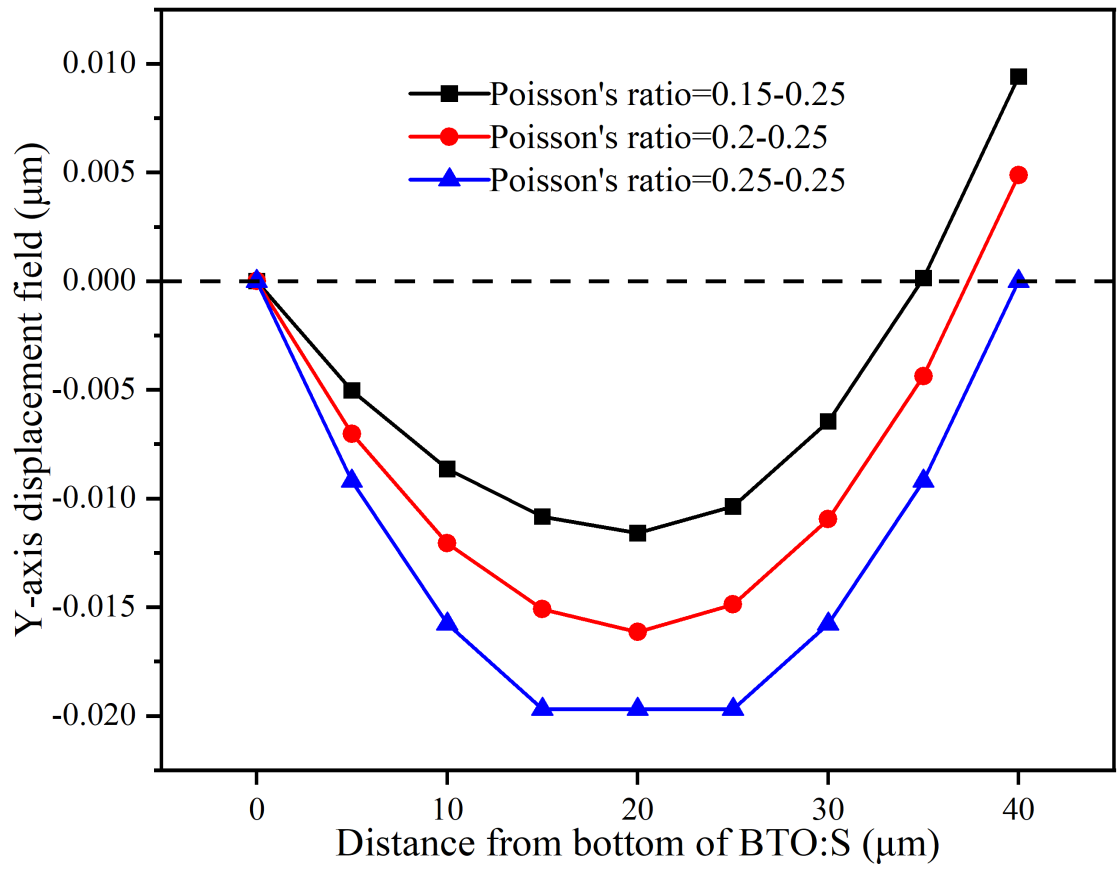

FIG. S6. Variation of Y-axis displacement field components with different membranes positions for different Poisson's ratio of nanofiber layers.

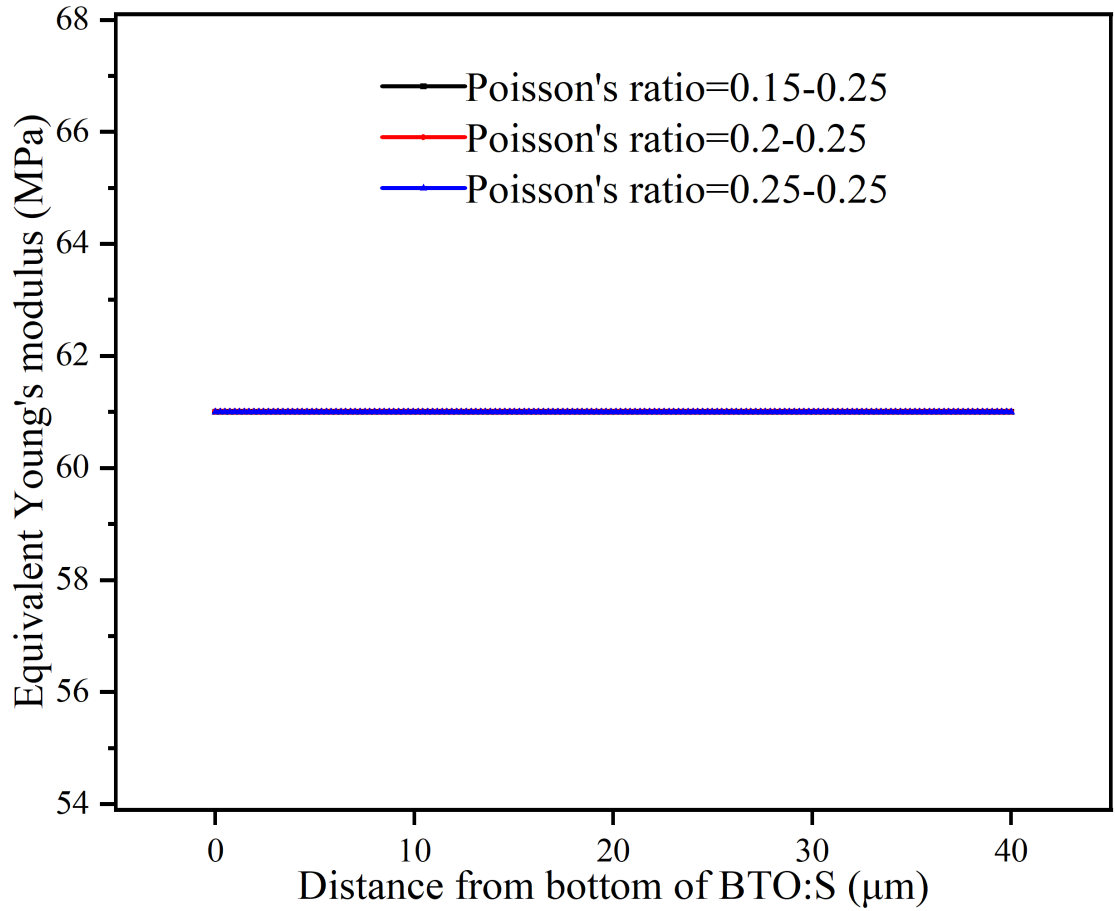

FIG. S7. Variation of Equivalent Young's modulus with different membranes positions at different Poisson's ratio of nanofiber layers.

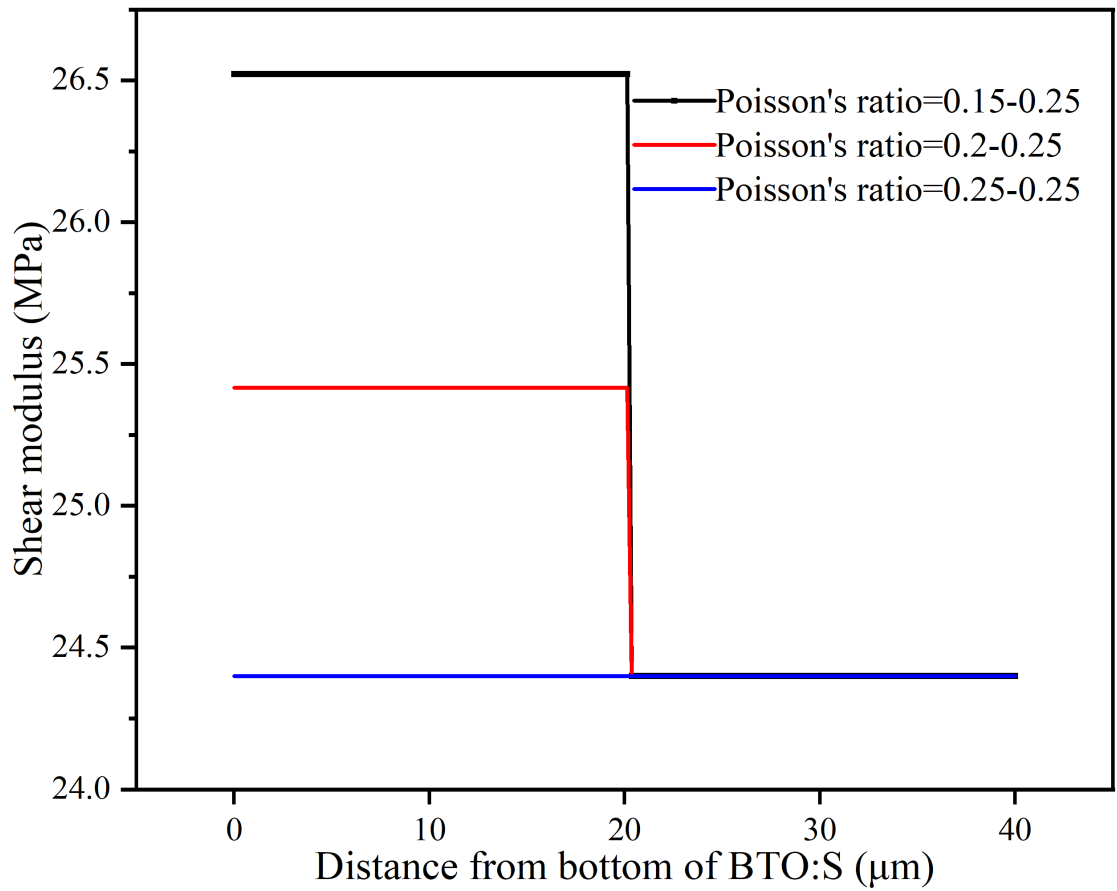

FIG. S8. Variation of shear modulus with different membranes positions at different Poisson's ratio of nanofiber layers.

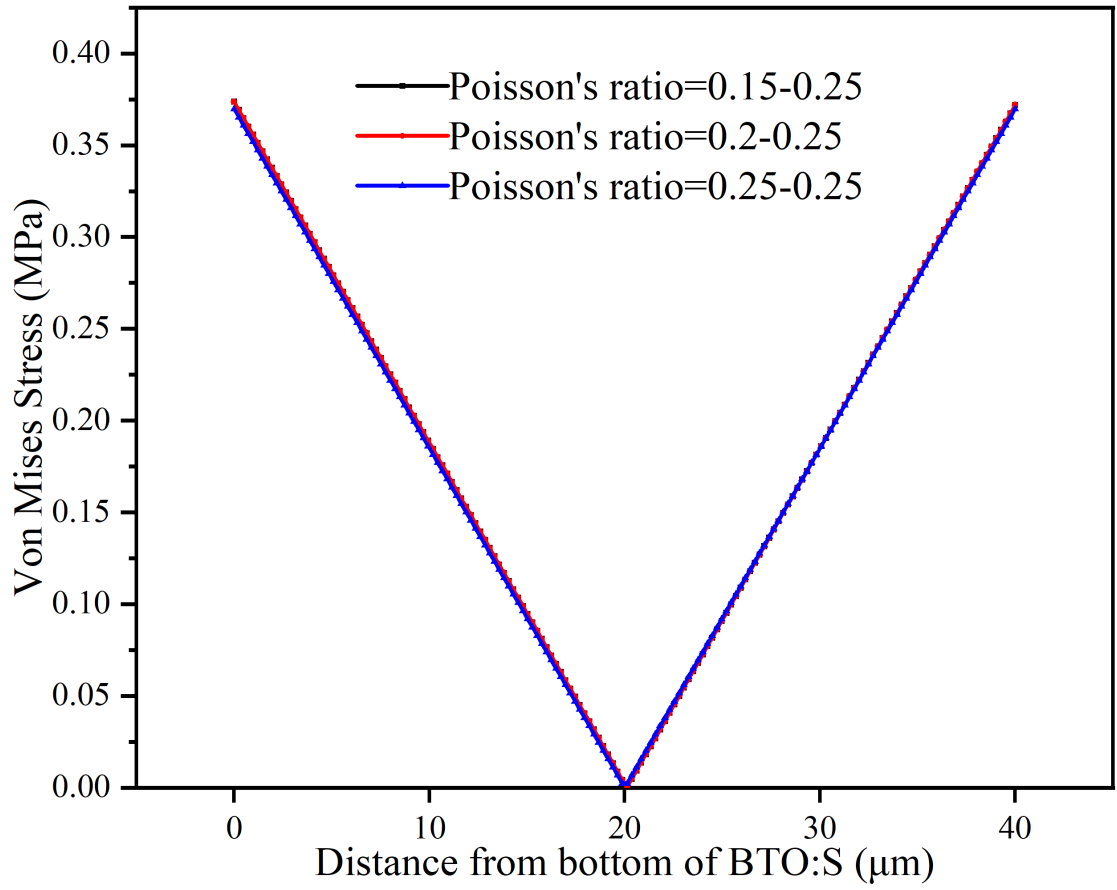

FIG. S9. Variation of von Mises stress with different membranes positions for different Poisson's ratio of nanofiber layers.

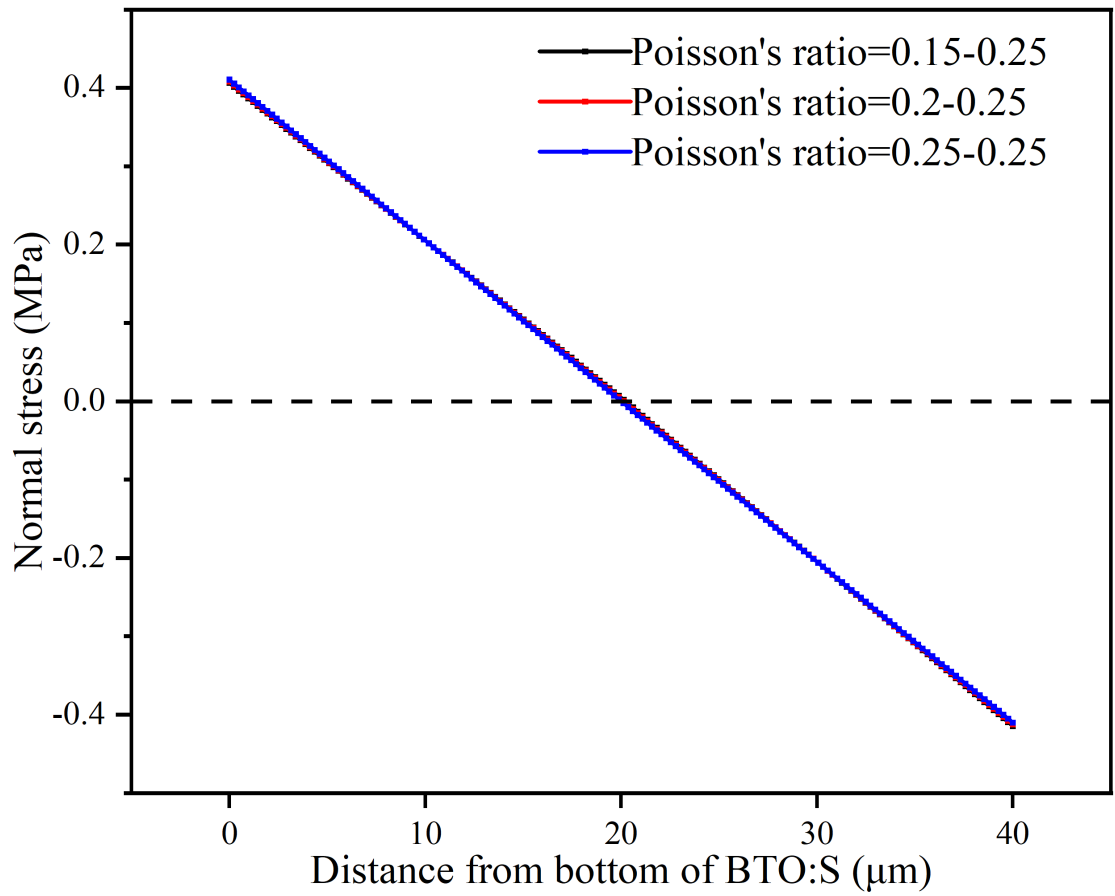

FIG. S10. Variation of normal stresses with different membranes positions at different Poisson's ratio of nanofiber layers.

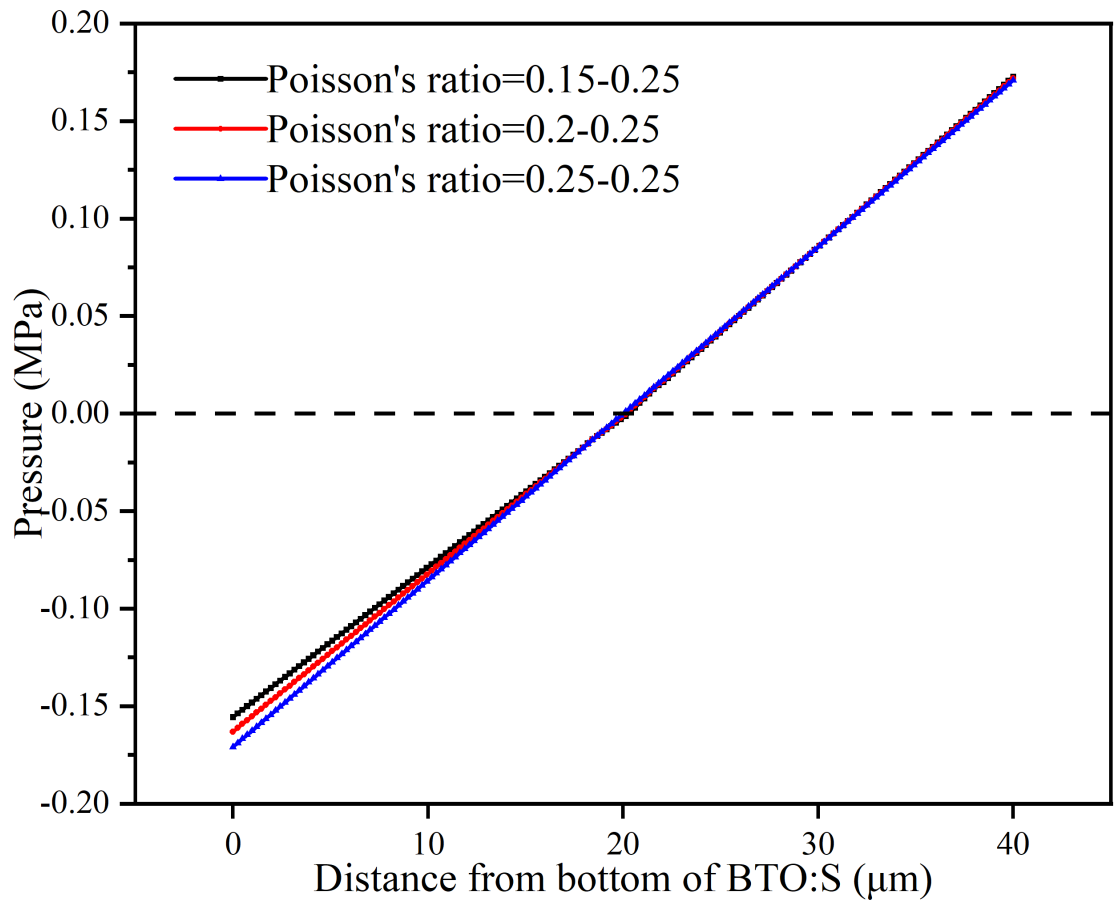

FIG. S11. Pressure variation with membranes positions at different Poisson's ratio of nanofiber layers.

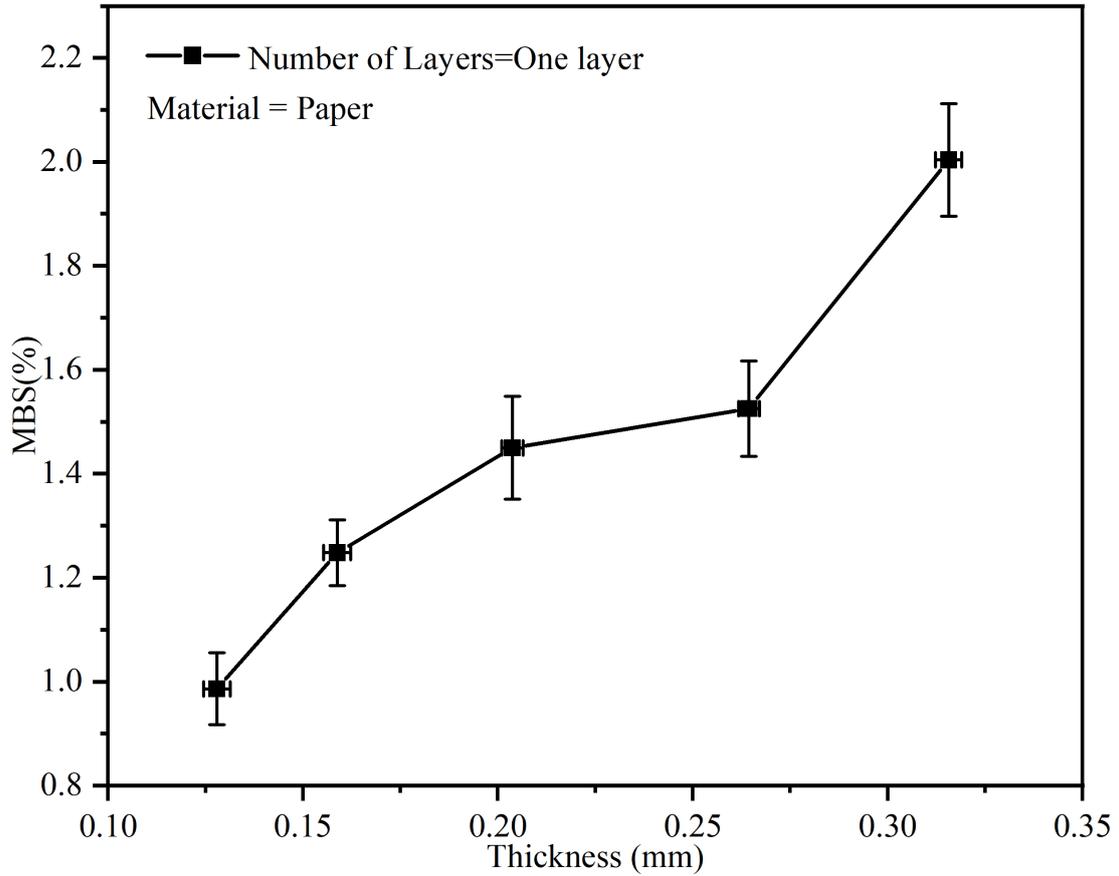

FIG. S12. In the case of a single layer of paper, the relationship between the MBS of the paper and the thickness.

In the experimental test, we tested the mechanical properties of paper sheets with different thicknesses for 20 times by a three-point bending test instrument. The test results are shown in Fig. S12. It can be seen that in the case of a single layer, the MBS of the paper sheet increases with the increase of thickness. All papers are the same batch of products produced by Youtai flagship store manufacturers using the same process. The distance between the squares of the paper sheet is 10 mm. The downward bending distance of the paper is obtained by the downward pressure distance of the probe, and the downward pressure rate of the probe is 0.01 mm·s$^{-1}$., and the original data is obtained by the change of the force of the paper sheet on the probe through the commercial balance. Then, the bending stress-strain relationship of the material is obtained by processing the original data. The original data and processed data in the experiment are sorted and placed in a compressed file.

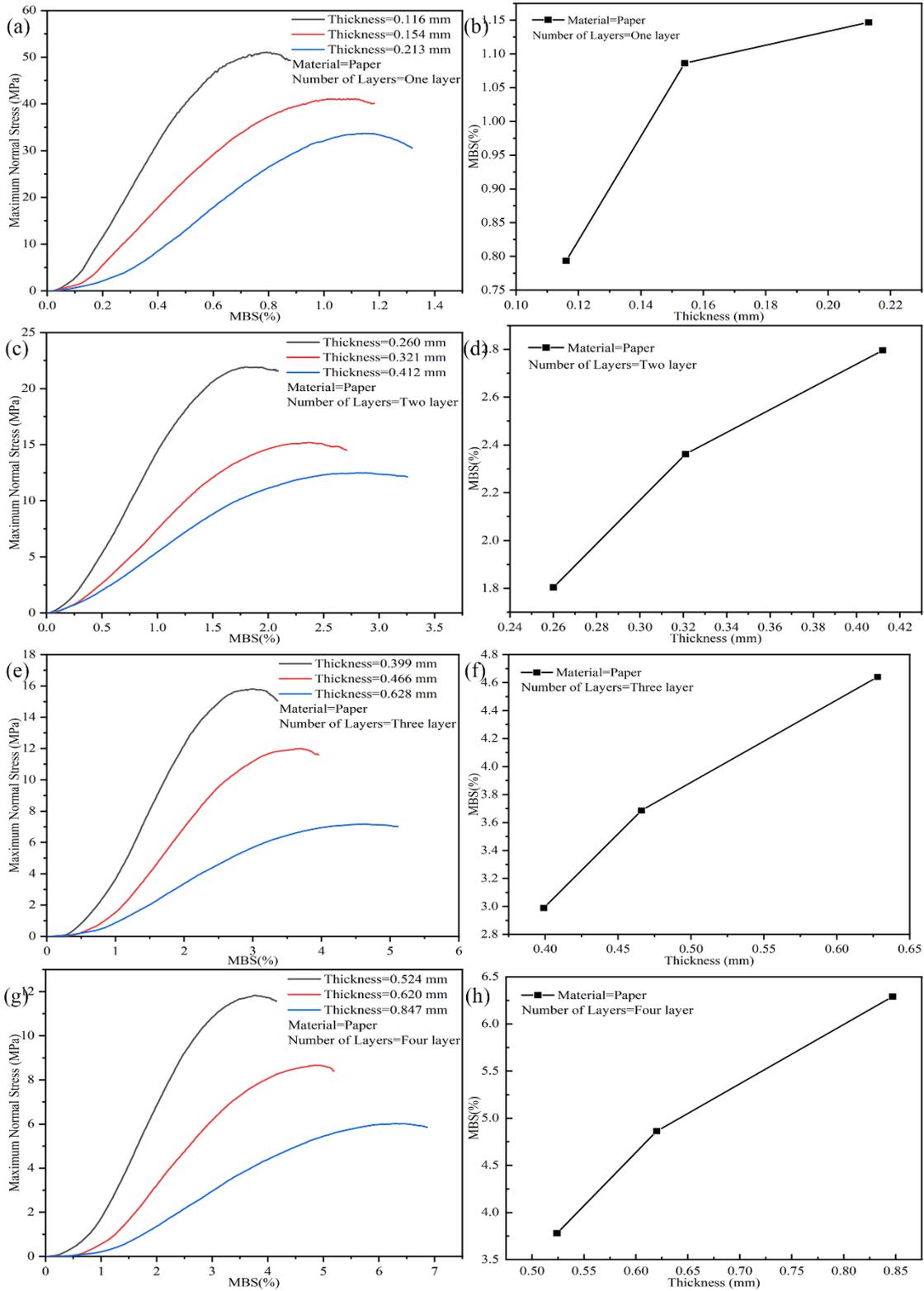

FIG. S13. In the case of different layers of paper, the mechanical properties of the paper change with the total thickness of the paper. Fig. (a), (c), (e) and (g) are the stress-strain relationship of paper under different layers. Fig. (b), (d), (f) and (h) are the relationship

between the maximum bending strain of the paper and the thickness corresponding to the stress-strain relationship in the left diagram, respectively.

Further in the experiment, we measured the mechanical properties of the paper at different layers of the paper. It can be seen from Fig. (b), (d), (f) and (h) that the MBS of the paper increases with the increase of the thickness in the case of different layers. The paper used in the experiment is still A4 paper made by the flagship store manufacturer. The experimental test methods and conditions are consistent with the test of Fig. S12.

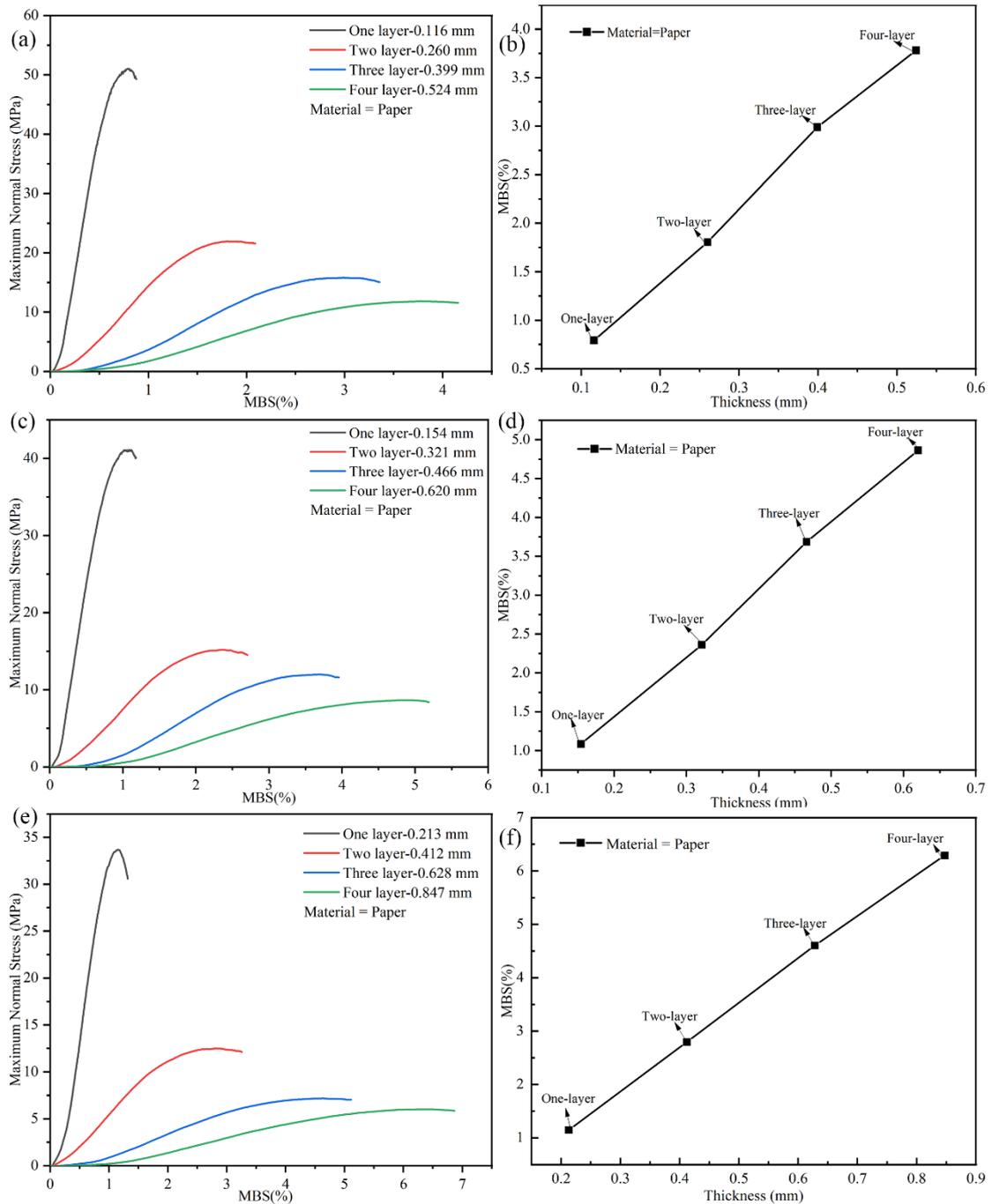

FIG. S14. The relationship between the mechanical properties of the paper under the same thickness of the paper and the number of paper layers. Fig. (a), (c) and (e) is the stress-strain diagram of paper in different layers. Fig. (b), (d) and (f) are the relationship between the MBS of the paper and the thickness corresponding to the stress-strain relationship in the left diagram, respectively.

In the experiment, the paper of the same A4 paper was cut to be laminated for testing. The mechanical properties obtained are shown in Figure S14. From Fig. (b), (d) and (f), it can be seen that as the number of layers is superimposed, the MBS of the paper increases with the increase of thickness. Fig. S13 and Fig. S14 use the same experimental data. The original data measured by the experiment and the processed data are collated and placed in a compressed file. In addition, the mechanical properties of Pb(Zr,Ti)O$_3$ (PZT), Seaweed sheet and Wall skin under different layers were tested. All of them showed a tendency for the MBS to increase with the increase in thickness, and all of the original and processed data were organized and placed in a compressed file.